\documentclass[showpacs,twocolumn,showkeys,amsmath,amssymb,pra]{revtex4} 

\pdfoutput=1

\usepackage{graphicx}   
\usepackage{dcolumn}    
\usepackage{bm}         
\newcommand{\rd}{{\rm d}} 
 
\newcommand{\f}{a}
\newcommand{\xx}{h}

\newcommand{\ri}{{i}}

\begin{document} 
 
\title{Fractional photon-assisted tunneling for Bose-Einstein condensates in a double well} 

\author{Niklas Teichmann}
\author{Martin Esmann}
\author{Christoph Weiss}
\email{christoph.weiss@uni-oldenburg.de}

\affiliation{Institut f\"ur Physik, Carl von Ossietzky Universit\"at,
                D-26111 Oldenburg, Germany
}

                  
\date{\today}
 
\begin{abstract}
Half-integer photon-resonances in a periodically shaken double well are investigated
on the level of the $N$-particle quantum dynamics. Contrary to non-linear mean-field
equations, the \emph{linear}
$N$-particle Schr\"odinger equation does not contain any non-linearity which could be the
origin of
such resonances.
Nevertheless, analytic calculations on the $N$-particle level
explain why such resonances can be observed even for particle numbers as low as $N=2$. These
calculations also demonstrate why fractional photon resonances are not restricted to
half-integer values.
\end{abstract} 
\maketitle


\section{Introduction}
Tunneling control of ultra-cold atoms via time-periodic
shaking~\nocite{GrossmannEtAl91,Holthaus92}\nocite{EckardtEtAl05}\cite{GrossmannEtAl91,Holthaus92,EckardtEtAl05,CreffieldSols08} of potentials is currently established
as an experimental method both on the single particle level~\cite{KierigEtAl08} and on the level of Bose-Einstein
condensates (BECs)~\cite{Zenesini09}. 
An interesting effect is an analog of photon-assisted tunneling in periodically shaken
systems of ultra-cold atoms. It was predicted theoretically both for the case that the
driving frequency matches the potential difference between neighboring
wells~\cite{KohlerSols03,EckardtEtAl05} and for the case that the driving frequency is
resonant with the interaction energy~\cite{CreffieldMonteiro06}. 
The $n$-photon resonances essentially are a single-particle
effect which survives interactions; one- and two-photon resonances have been observed
experimentally for BECs in periodically shaken lattices~\cite{SiasEtAl08}. The ``photons''
are time-periodic potential modulations in the kilo-Hertz-regime.

However, photon-assisted
tunneling is not restricted to integer photon resonances. Also
half-integer Shapiro-like~\cite{Shapiro63} resonances have been predicted numerically both on the mean-field
(Gross-Pitaevskii) level and on the level of the multi-particle quantum dynamics (down to
$N=2$ particles)~\cite{EckardtEtAl05}. 
While the occurrence of higher or lower harmonics in non-linear equations
is easy to understand qualitatively, it is not
clear a priori how these resonances should occur in the linear $N$-particle
Schr\"odinger equation. Thus,  analytic calculations which can explain the occurrence of such
resonances within the linear quantum dynamics will explain how effective non-linearities can arise from linear
dynamics even for small particle numbers. Realistic experimental values for the number of atoms in a double well can be of the order of
1000 atoms~\cite{AlbiezEtAl05} for BECs and down to less than 6 atoms~\cite{CheinetEtAl08} for
few-atoms experiments.

Often, Floquet-states~\cite{Shirley65} help to understand the physics of BECs in periodically
driven systems~\cite{JinasunderaEtAl06,EckardtHolthaus08b,StrzysEtAl08,HaiEtAl08}. The focus
of the present
paper, lies on a different approach: analytic calculations on the $N$-particle
level developed in Ref.~\cite{WeissJinasundera05} (cf.\ Ref.~\cite{KalosakasEtAl03}). By assuming the
experimentally realistic initial condition of all particles being in one
well~\cite{AlbiezEtAl05}, the calculations are done analogously to the time-dependent perturbation theory.

The paper is organized as follows: after introducing the two-mode model for a BEC in a double
well (Sec.~\ref{sec:a}), we develop the technique to calculate half-integer resonances in
Sec.~\ref{sec:the}. A crucial test is to show that the analytic result vanishes in the limit
of non-interacting particles (Sec.~\ref{sec:half}). Other fractional resonances are discussed in Sec.~\ref{sec:frac}.

\section{\label{sec:a}The Model: a BEC in a double well}
Bose-Einstein condensates in double-well potentials are interesting both
experimentally and theoretically~\nocite{SmerziEtAl97}\nocite{CastinDalibard97}\nocite{LesanovskyEtAl06}\cite{AlbiezEtAl05,SmerziEtAl97,CastinDalibard97,LesanovskyEtAl06,PiazzaEtAl2008,LeeEtAl08,EsteveEtAl08,YukalovYukalova09}.
In order to describe a BEC in a double well, we use a model
originally developed in nuclear physics~\cite{LipkinEtAl65}: a multi-particle Hamiltonian in two-mode
approximation~\cite{MilburnEtAl97}, 
\begin{eqnarray}
\label{eq:H}
\hat{H} &=& -\frac{\hbar\Omega}2\left(\hat{c}_1^{\phantom\dag}\hat{c}_2^{\dag}+\hat{c}_1^{\dag}\hat{c}_2^{\phantom\dag} \right) + \hbar\kappa\left(\hat{c}_1^{\dag}\hat{c}_1^{\dag}\hat{c}_1^{\phantom\dag}\hat{c}_1^{\phantom\dag}+\hat{c}_2^{\dag}\hat{c}_2^{\dag}\hat{c}_2^{\phantom\dag}\hat{c}_2^{\phantom\dag}\right)\nonumber\\
&+&\hbar\big(\mu_0+\mu_1\sin(\omega t)\big)\left(\hat{c}_2^{\dag}\hat{c}_2^{\phantom\dag}-\hat{c}_1^{\dag}\hat{c}_1^{\phantom\dag}\right)\;,
\end{eqnarray}
where the operator $\hat{c}^{(\dag)}_j$ annihilates (creates) a boson in well~$j$;
$\hbar\Omega$ is the tunneling splitting, $\hbar\mu_0$ is
the tilt between well~1 and well~2 and $\hbar\mu_1$ is the driving amplitude. The interaction
between a pair of particles in the same well is denoted by $2\hbar\kappa$.

The Gross-Pitaevskii dynamics can be mapped to that  of a nonrigid
pendulum~\cite{SmerziEtAl97}. Including the term  describing the  periodic shaking, the 
Hamilton function is given by:
\begin{eqnarray}
\label{eq:mean}
H_{\rm mf}& = &\frac{N\kappa}{\Omega}z^2-\sqrt{1-z^2}\cos(\phi)\nonumber\\
&-&2z\left(\frac{\mu_0}{\Omega}+
\frac{\mu_1}{\Omega}\sin\left({\textstyle\frac{\omega}{\Omega}}\tau\right)\right)\;,\quad \tau =t\Omega\;,
\end{eqnarray}
where $\phi$ and $z$ are canonically conjugate variables. 
The quantity  $z/2$ is the population imbalance with $z/2=0.5$ ($z/2=-0.5$) referring to the situation with all
particles in well~1 (well~2). The corresponding observable on the $N$-particle level is given
by
\begin{equation}
\frac{J_z(t)}{N}=\frac{\langle\Psi(t)|\hat{c}_1^{\dag}\hat{c}_1^{\phantom\dag}-\hat{c}_2^{\dag}\hat{c}_2^{\phantom\dag}|\Psi(t)\rangle}{2N}.
\end{equation}


For integer photon-assisted tunneling, the potential difference between both wells, $2\hbar\mu_0$
has to be bridged by an integer number of photons:
\begin{equation}
\label{eq:integerphoton}
2\hbar\mu_0 = n\hbar\omega\;, \quad n=1,2,\ldots
\end{equation}
The $1/2$-integer resonance occurs for
\begin{equation}
\label{eq:halfphoton}
2\hbar\mu_0 = \frac 12\hbar\omega\;;
\end{equation}
for an interacting bose gas, these resonances are furthermore shifted~\cite{EckardtEtAl05}.

For some parameter regimes (especially for interactions comparable to the onset of the
self-trapping transition~\cite{AlbiezEtAl05}), the differences between mean-field
(Gross-Pitaevskii) dynamics and the $N$-particle quantum dynamics can be quite
remarkable~\cite{WeissTeichmann08}. However, when concentrating on the (experimentally
measurable~\cite{AlbiezEtAl05}) time-averaged population imbalance,
\begin{equation}
\label{eq:timeaveraged}
\frac{\langle J_z\rangle_T}N=\frac1{T}\int_0^T\frac{J_z(t)}N dt,
\end{equation}
 for \emph{low}
interactions, the qualitative agreement between mean-field and $N$-particle dynamics for the occurrence of both integer and half-integer
photon-assisted tunneling is excellent~\cite{EckardtEtAl05}.

\begin{figure}
\includegraphics[width=0.6\linewidth,angle=-90]{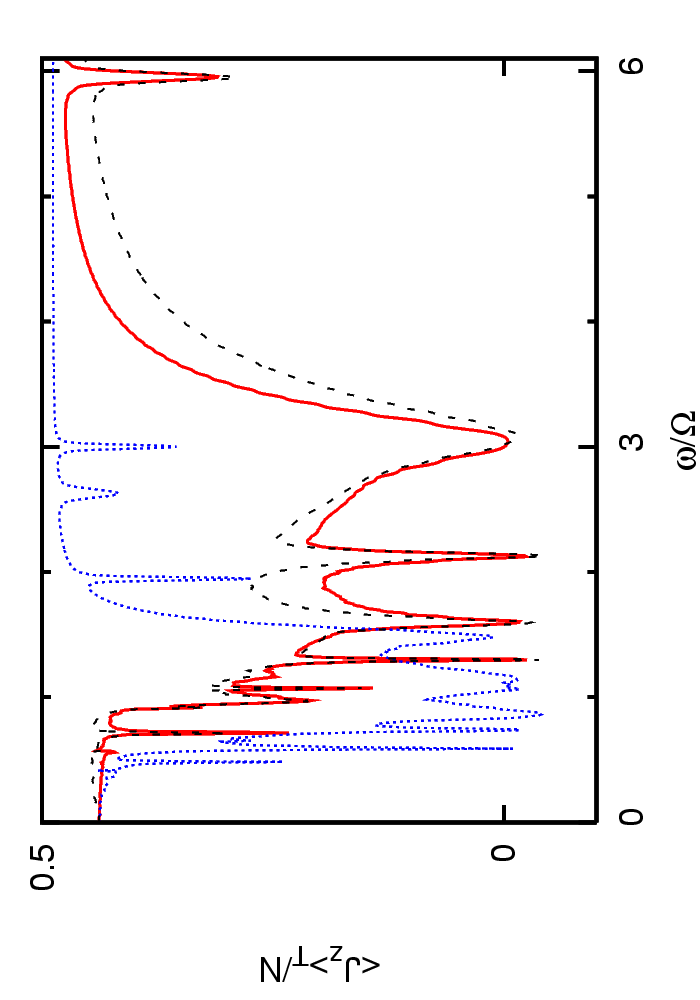}
\caption{\label{fig:quasi}
(Color online) Time-averaged population imbalance $\langle J_z \rangle_T/N$ with
averaging time $T=100/\Omega$, static tilt $2\mu_0/\Omega = 3$ and
interaction $N\kappa/\Omega=0.4$ for $N=2$ particles initially in well~1. The red
(full) line with driving amplitude $2\mu_1/\omega = 1.8$ (close to the
maximum of the $J_1$-Besselfunction) has, among others,  a pronounced resonance at the
driving frequency $\omega = 3\Omega$ (one-photon resonance) and also a weaker resonance at $\omega
= 6\Omega$ (1/2-photon resonance).
As the black (dashed) line shows, these resonances are also visible when the system
is driven with an initial $\pi/2$ phase shift ($\mu_1\cos (\omega t)$
instead of $\mu_1\sin (\omega t)$ in Eq.~(\ref{eq:H})), while they are
strongly suppressed with driving amplitude $2\mu_1/\omega = 3.83$ (close to
the first zero of $J_1$) illustrated by the blue (dotted) line.
}
\end{figure}

Photon-assisted tunneling is clearly visible in the experimentally measurable time-averaged
population imbalance~(\ref{eq:timeaveraged}). Figure~\ref{fig:quasi} shows integer
resonances~(\ref{eq:integerphoton}), namely 
the one-photon peak with $\omega \approx 3\Omega$ and the two-photon peak at $\omega \approx
1.5\Omega$. Furthermore, there are pronounced fractional-integer resonances at $\omega\approx 6\Omega$,
$\omega\approx 2\Omega$ and $\omega\approx 1.2\Omega$ corresponding to the 1/2, 3/2 and
5/2-photon peaks. While some of the resonances disappear~\cite{EckardtEtAl05}
for specific choices of the driving \emph{amplitude}, the initial phase of the periodic
driving (cf.~\cite{CreffieldSols08}) does not influence the occurrence of resonances in the
situation investigated in this manuscript.

\begin{figure}
\includegraphics[angle=-90,width=0.8\linewidth]{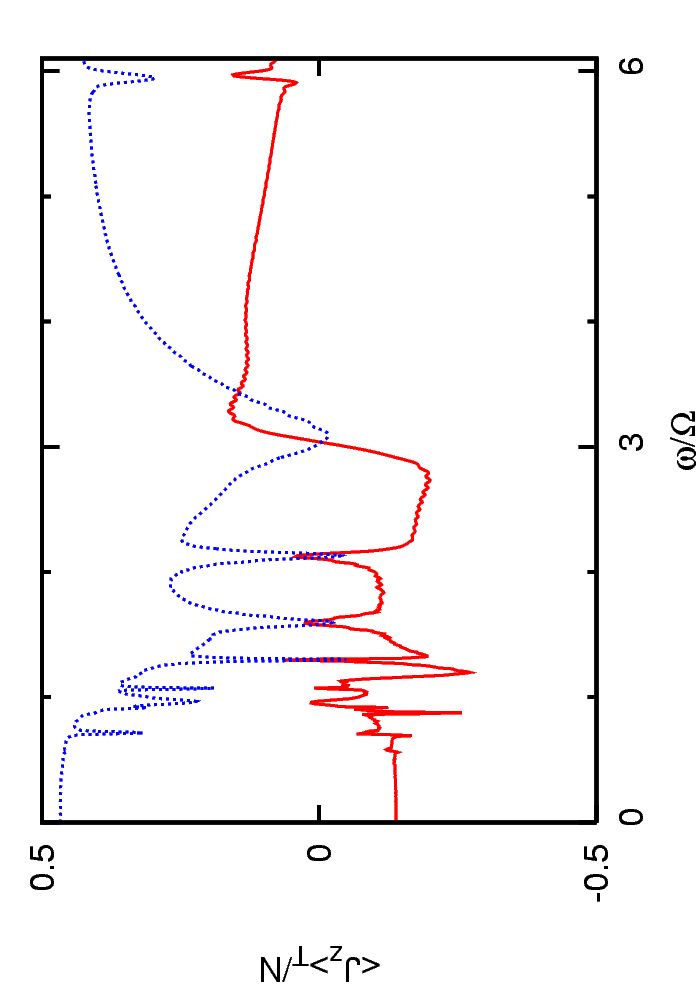}
\caption[test]{\label{fig:timeaveraged}(Color online) 
Time-averaged population imbalance $\langle J_z \rangle_T/N$ with
averaging time $T=100/\Omega$, static tilt $2\mu_0/\Omega = 3$ and
interaction $N\kappa/\Omega=0.4$ with $N=2$ particles for two different
initial states. The dotted line corresponds to the ground state of the
tilted system, while the full line displays $\langle J_z \rangle_T/N$
with the ground state of the untilted system as initial state. In both
cases the 1/2-photon resonance at $\omega = 6\Omega$ appears.
}
\end{figure}

Figure~\ref{fig:timeaveraged} shows that it is not essential to start with all particles in
one well in order to observe photon-assisted tunneling. Both for the ground-state of the
untilted, undriven system (for which the initial population imbalance is zero) and for the
ground-state of the tilted system with an initial population imbalance of $\simeq 0.467$,
the main resonances of Fig.~\ref{fig:quasi}, where all particles were initially in well~1, can easily be identified.

\begin{figure}
\vspace*{-1.5cm}

 \hspace*{-1cm}
\includegraphics[width=0.8\linewidth,angle=-90]{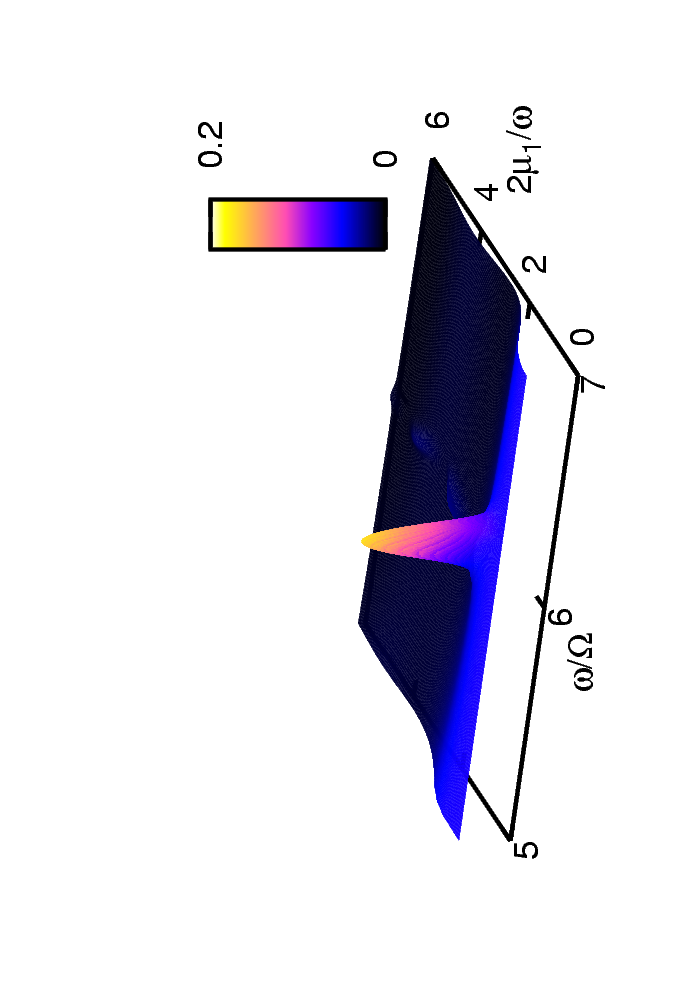}

\hspace*{-1cm}\includegraphics[width=0.8\linewidth,angle=-90]{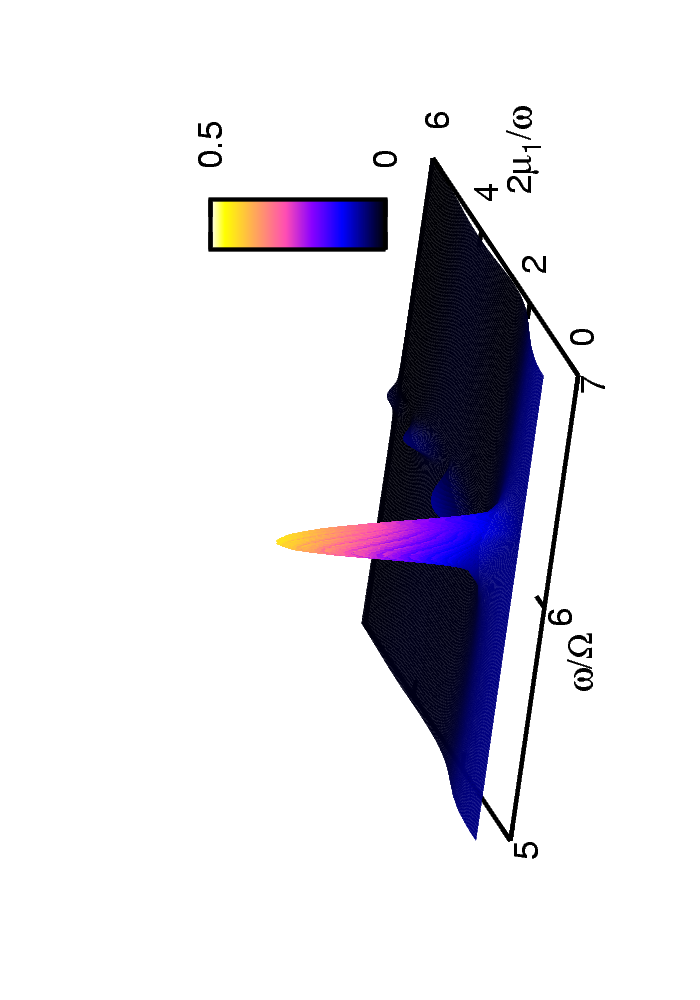}
\caption{\label{fig:einhalb}(Color online) Time-averaged population imbalance for $N=2$ particles in a
 periodically shaken double-well potential as a function of both driving frequency~$\omega$
 and driving amplitude~$\mu_1$ for a static tilt of $2\mu_0/\Omega = 3$. Upper panel: $N\kappa/\Omega=0.2$; 
 lower panel:
 $N\kappa/\Omega=0.4$. The averaging time is $\Omega T=100$; the values plotted are shifted
 such that 0 corresponds to all particles having always stayed in the first well (a value of 1 would
 correspond to all particles being in well 2).
 Surprisingly, contrary to the case of $N> 2$ or the mean-field case~\cite{EckardtEtAl05}, for $N=2$ particles the
  resonance does not shift with increasing interactions.}
\end{figure}

Figure~\ref{fig:einhalb} displays the half-integer resonance for $N=2$ particles. Contrary to
what was observed for both larger particle numbers and for mean-field, the position of the
$1/2$ photon resonance does not shift with increasing energy. A first test of our analytic
calculations towards the end of the next section will thus be to explain this feature.

\section{\label{sec:the}Analytic calculations}
In order to analytically describe the 
time-evolution of the interacting system, the 
Fock basis
$
   |\nu\rangle \equiv |N-\nu,\nu\rangle 
$   
is used. The label   
$   
   \nu=0\ldots N
$
refers to a state with $N-\nu$~particles in well~$1$,
and $\nu$~particles in well~$2$. The Hamiltonian~(\ref{eq:H}) now is the 
sum of two $(N+1)\times(N+1)$-matrices,
\begin{equation}
\label{eq:Hsum}
   H = H_0(t) + H_1 \; .
\end{equation}
While the non-diagonal matrix~$H_1$ is given by the tunneling-terms of Eq.~(\ref{eq:H}), 
the diagonal matrix~$H_0$ includes both the interaction between 
the particles and the applied potential difference.
For the solution of the Schr\"odinger equation
\begin{equation}
   \ri\hbar\frac{\partial}{\partial t}|\psi(t)\rangle = 
   \left(H_0(t)+H_1\right)|\psi(t)\rangle \; ,
\end{equation}
the ansatz 
\begin{equation}
   \langle \nu|\psi(t)\rangle = 
   \f_\nu(t)\exp\left[-\frac{\ri}{\hbar}\int_0^t
   \langle \nu | H_0(t')|\nu\rangle\rd t'\right] \; 
\end{equation}
turned out to be useful~\cite{WeissJinasundera05}.

Within this framework, a 
set of differential equations was derived~\cite{WeissJinasundera05} which is mathematically equivalent to the
$N$-particle Schr\"odinger equation governed by the Hamiltonian~(\ref{eq:H}):
\begin{eqnarray}
\label{eq:equiv}
   \ri\hbar\dot{\f}_{\nu}(t) & = &
   \langle\nu | H_1 |\nu\!+\!1\rangle\xx_{\nu}(t){\f}_{\nu+1}(t)
\nonumber\\
   & + & \langle\nu|H_1|\nu\!-\!1\rangle \xx_{\nu-1}(t)^*{\f}_{\nu-1}(t) \;.
\end{eqnarray}
In Eq.~(\ref{eq:equiv}), the notation
$
   a_{-1}(t) \equiv a_{N+1}(t) \equiv 0 \, ,
$
was used; the phase factors are given by:
\begin{equation}
\label{eq:phase}
   \xx_{\nu}(t) = 
\exp\left({\textstyle \ri\left[2(N - 1 - 2\nu)\kappa t 
   +2\mu_0t-2\mu_1\cos(\omega t)/\omega\right] }\right)
\end{equation}
with $-\cos(\omega t)/\omega = \int_{0}^t\sin(\omega t') dt' $. To simplify the expression
for subsequent integrals, one can use the expansion in terms of Bessel functions~\cite{Abramowitz84}
\begin{equation}
\label{eq:bessel}
e^{iz\cos(\omega t)}=\sum_{k=-\infty}^{\infty}J_{k}(z)i^ke^{ik\omega t}\;.
\end{equation}
Equation~(\ref{eq:equiv})
furthermore needs
\begin{eqnarray}
\langle\nu | H_1 |n\rangle =
&-&\frac {\hbar\Omega}{2}\delta_{\nu,n+1}\sqrt{N-n}\sqrt{n+1}\\ \nonumber
&-&\frac {\hbar\Omega}{2}
\delta_{\nu,n-1} \sqrt{N-n+1}\sqrt{n}
 \;,
\end{eqnarray}
where $\delta_{n,m}$ is the Kronecker delta (which is zero except for $n=m$ where  $\delta_{n,n}=1$).
The idea is to proceed along the lines of time-dependent perturbation theory~\cite{Landau00}.
Starting with a typical experimental initial condition such that all particles are in the
first
well~\cite{AlbiezEtAl05}, one has in zeroth order perturbation theory:
\begin{equation}
 a_0^{(0)}(t)= 1\,, \quad a_1^{(0)}(t)= a_2^{(0)}(t)=\ldots=0\;,
\end{equation}
where $a_{\nu}= \sum_{k=0}^{\infty}a_{\nu}^{(k)}$
In first order perturbation theory, one gets:
\begin{equation}
 a_{\nu}^{(1)}(t)=0
\end{equation}
if  $\nu\ne1$ and
\begin{equation}
 a_1^{(1)}(t) = i \frac{\Omega}2\sqrt{N}\int_{0}^{t}h_0^*(t)a_0^{(0)}(t)\;.
\end{equation}
Using Eqs.~(\ref{eq:phase}) and (\ref{eq:bessel}) one thus has
\begin{equation}
a_1^{(1)}(t) = i
\frac{\Omega}2\sqrt{N}\sum_{k=-\infty}^{\infty}i^kJ_k(2\mu_1/\omega)\int_{0}^{t}\exp\left(i\sigma_kt'\right)dt'
\label{eq:a11}
\end{equation}
with
\begin{equation}
\sigma_k\equiv k\omega -2\mu_0-2(N-1)\kappa\;.
\end{equation}
Therefore, after solving the integral, Eq.~(\ref{eq:a11}) is a sum of time-periodic functions
except for the special case with $\sigma_k=0$ which recovers the integer photon resonances of
Eq.~(\ref{eq:integerphoton})
investigated in Refs.~\cite{EckardtEtAl05,SiasEtAl08}. While for a double
well as in Ref.~\cite{EckardtEtAl05}, the population imbalance is ideal to investigate photon-assisted
tunneling,  the experiment~\cite{SiasEtAl08} was performed in an optical lattice. The
signatures of photon-assisted tunneling were seen in the width of the
BEC after expansion in the shaken lattice. Surprisingly,  a
$J_n^2$ dependence was measured. While this might be interpreted as being an indication for transition
from ballistic to diffusive transport~\cite{SiasEtAl08}, the present experiments cannot
exclude other explanations. The $J_n^2$-dependence could either be an interaction-induced
effect~\cite{WeissBreuer09} or the result of an effective average over the precise
instant within the cycle at which the current is measured~\cite{CreffieldSols08}.

As the aim of the present paper is to
understand the fractional photon peaks like the interaction induced half-integer resonances
of Ref.~\cite{EckardtEtAl05}, we can discard the integer-photon resonances characterized via
$\sigma_k=0$~\footnote{For $\sigma_k=0$ the amplitude $a_1^{(1)}$ would contain a part which
  increases linearly in time. This both signifies the break-down of our perturbation
theory (for too large times) and the onset of photon-assisted tunneling.} and thus write
\begin{equation}
a_1^{(1)}(t) = i
\frac{\Omega}2\sqrt{N}\sum_{k=-\infty}^{\infty}i^kJ_k(2\mu_1/\omega)\frac{e^{i\sigma_kt}-1}{i\sigma_k}
\label{eq:a11end}
\end{equation}
In second order perturbation theory one has (see the appendix~\ref{sec:second}):
\begin{eqnarray}
\label{eq:textsecond}
a_2^{(2)}(t) &=&\left(i\frac{\Omega}2\right)^2\sqrt{N}\sqrt{N-1}\sqrt{2}\\\nonumber
&\times&\sum_{k=-\infty}^{\infty}\sum_{\ell=-\infty}^{\infty}i^ki^{\ell}J_k(2\mu_1/\omega)J_{\ell}(2\mu_1/\omega)\\\nonumber
&\times&\int_0^t\frac{\exp(i\sigma_kt')-1}{i\sigma_k}\exp(i\widetilde{\sigma}_{\ell}t')dt'
\end{eqnarray}
with
\begin{equation}
\widetilde{\sigma}_{\ell}\equiv \ell\omega -2\mu_0-2(N-3)\kappa\;.
\end{equation}
Again, $\widetilde{\sigma}_{\ell}=0$ can be discarded because it corresponds to integer photon
resonances. However, if $\widetilde{\sigma}_{\ell}+\sigma_k=0$ then $a_2^{(2)}$ does have parts
which increase linearly in time. In order to see that this indeed corresponds to a
half-integer resonance, we choose $N=2$ and
$\omega/2 =2\mu_0$. This implies $\sigma_k = k\omega-\omega/2-2\kappa$ and 
$\widetilde{\sigma}_{\ell} =\ell \omega -\omega/2+2\kappa$ and the condition 
\begin{equation}
\label{eq:summesigma}
\widetilde{\sigma}_{\ell}+\sigma_k=0
\end{equation}
 thus
becomes independent of the interaction; it results in the simple equation 
\begin{equation}
\label{eq:summekell}
k+\ell =1\;.
\end{equation} 

The above reasoning explains why we observe no shift of the resonance with increasing
interaction in the numerics displayed in
Fig.~\ref{fig:einhalb}. The
amplitude to find both particles in well two is given by (appendix~\ref{sec:second}):
\begin{eqnarray}
\label{eq:a22}
a_2^{(2)}(t) = \left(a_2^{(2)}(t)\right)_{\rm oscil}\quad\quad \quad  \\\nonumber-\frac{\Omega^2}2
\sum_{k=-\infty}^{\infty}J_k\left(2\mu_1/\omega\right)J_{1-k}\left(2\mu_1/\omega\right)\frac t{\sigma_k}\\\nonumber -\frac{\Omega^2}2
\sum_{k=-\infty}^{\infty}J_k\left(2\mu_1/\omega\right)J_{1-k}\left(2\mu_1/\omega\right)\frac
{e^{-i{\sigma}_{k}t}-1}{i\sigma_k}\;,
\end{eqnarray}
where the expression $(a_2^{(2)}(t))_{\rm oscil}$ contains oscillatory terms
which can be found in Eq.~(\ref{eq:a2oscil}). The convergence of this sum is ensured both
by the scaling of $\sigma_k$ on $k$ and the behavior of Bessel functions with increasing
$k$~\cite{Abramowitz84}
\begin{equation}
 J_k(z)\sim \frac1{\sqrt{2\pi k}}\left(\frac{z\exp(1)}{2k}\right)^k\;, \quad k\to\infty
\end{equation}
 combined with the fact that $J_{-k}(x)=(-1)^k J_k(x)$ (for integer $k$).
\begin{figure}
\hspace*{-1cm}\includegraphics[width=0.6\linewidth,angle=-90]{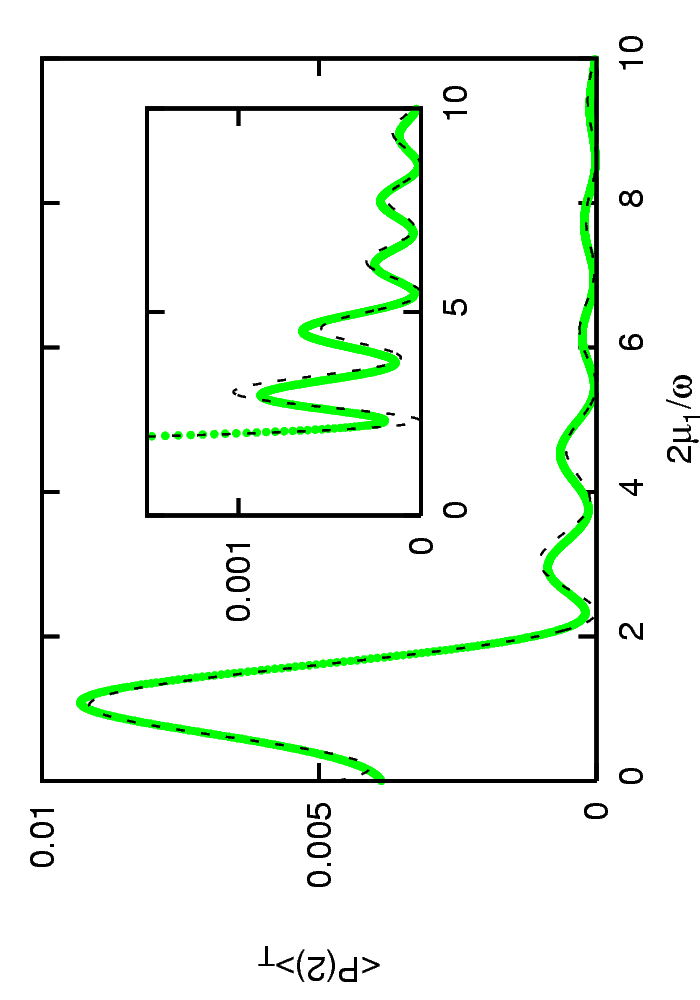}
\caption{\label{fig:a2}(Color online) Time-averaged probability (averaged over $T\Omega = 10$) that two particles have tunneled to the
  other well for the $1/2$-photon resonance ($N=2$, $\mu_0=3\Omega/2$, $\omega=6\Omega$, $\kappa=0.2\Omega$). Wide (green/grey) line: numerical data; dashed (black) line: perturbation theory
  (cf.\ Eq.~(\ref{eq:a22nah})). The
  probabilities displayed here should be measurable experimentally (see Ref.~\cite{CheinetEtAl08}).}
\end{figure}
Figure~\ref{fig:a2} shows good qualitative agreement between the analytic and
numeric calculations for the time-averaged probability that both particles, which initially
have been in the first well, have tunneled to
the second well. Already perturbation theory in the first order, in which the half-integer
resonance becomes visible, correctly describes the occurrence of maxima and minima in the
probability for both particles to occupy the second well.

\section{\label{sec:half}Half-integer resonances disappear in the limit of low interactions}
Despite the agreement displayed in Fig.~\ref{fig:a2}, at the first glance Eq.~(\ref{eq:a22}) seems to contain a flaw: numerically, we observe that the
half-integer resonance disappears for zero interaction. However, there seems to be a sum of non-zero
terms proportional to $t$ even for $\kappa = 0$. As it is not obvious that these terms cancel,
the next step will be to demonstrate that $a_2^{(2)}$ indeed approaches zero for vanishing
interaction.

As shown in the appendix, the terms proportional to $t$ in $a_2^{(2)}$ are due to
situations such that Eq.~(\ref{eq:summesigma}) is fulfilled. In the limit~$\kappa\to 0$ this
results again in the condition~(\ref{eq:summekell}), independent of the particle number. The
part of $a_2^{(2)}$ which increases linearly in time is thus
proportional to
\begin{equation}
A\equiv \sum_{k=-\infty}^{\infty}J_k(2\mu_1/\omega)J_{1-k}(2\mu_1/\omega)\frac1{\sigma_k}\;.
\end{equation}
Dividing the sum into two parts ($\sum_{k=1}^{\infty}\dots+\sum_{\ell=-\infty}^{0}\dots$) and
then setting $1-\ell = k$, one obtains
\begin{eqnarray}
A=\sum_{k=1}^{\infty}J_k(2\mu_1/\omega)J_{1-k}(2\mu_1/\omega)\left(\frac1{\sigma_k}+
\frac1{\sigma_{1-k}}\right)\;.
\end{eqnarray}
In the limit $\kappa\to 0$, the position of the half-integer resonance approaches the value
for $N=2$ particles. Therefore, one has $\sigma_k=k\omega -\omega/2$ and thus
$\sigma_{1-k}=-\sigma_k$ which implies
\begin{equation}
A=0\;.
\end{equation}
Thus, in agreement with the numerics, the half-integer photon peak disappears with vanishing
interactions.

\section{\label{sec:frac}Fractional integer resonances}
\begin{figure}
\includegraphics[angle=-90,width=1.1\linewidth]{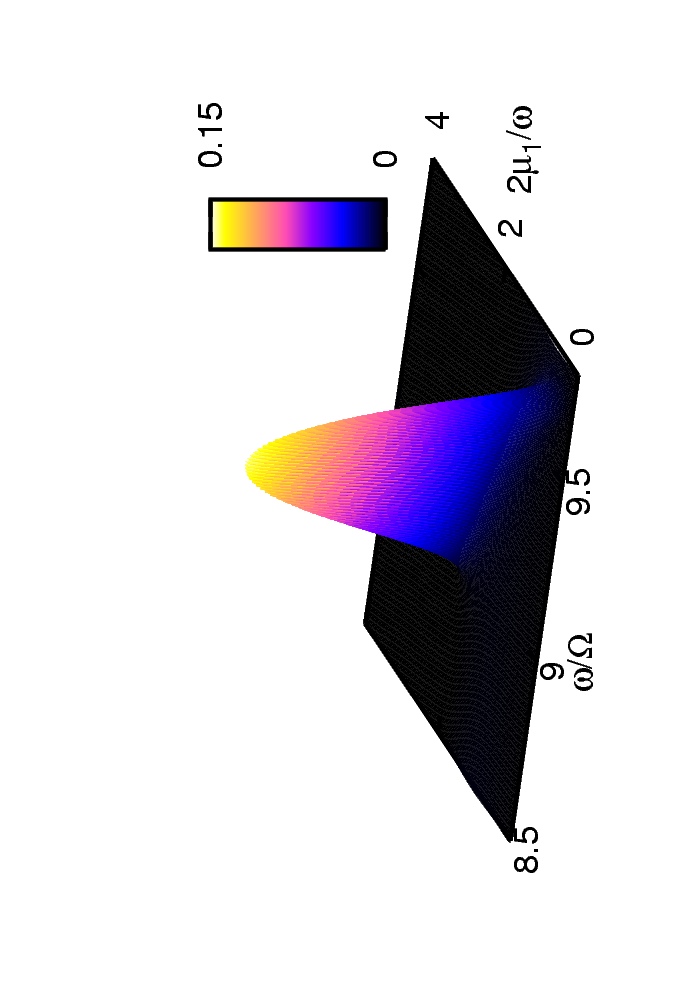}
\caption[]{\label{fig:drittel}The plot illustrates the
$1/3$-resonance for $N=3$ particles and interaction parameter
$N\kappa/\Omega = 1.5$. Initially, all three particles were in the lower well; the static
tilt is again given by $2\mu_0/\Omega = 3$. The time averaged probability (averaged over time
$T \Omega = 100$) to find all particles in the upper well as a function
of the driving frequency $\omega/\Omega$ and the driving amplitude
$2\mu_1/\omega$ has a clear peak at $\omega \approx 9 \Omega$.}
\end{figure}

Fractional integer resonances are not, however, restricted to the half-integer resonances
investigated numerically in Ref.~\cite{EckardtEtAl05} and analytically in
Sec.~\ref{sec:half}. For 
$N=3$ particles and a driving frequency such that $\omega/3=2\mu_0$, the condition
\begin{equation}
\sigma_k+\widetilde{\sigma}_{\ell}+ \widetilde{\widetilde{\sigma}}_{m} =0
\end{equation}
with 
\begin{equation}
\widetilde{\widetilde{\sigma}}_{m} =m\omega-2\mu_0-2(N-5)\kappa
\end{equation}
(throughout this section: $N=3$) is fulfilled for
\begin{equation}
k+\ell+m =1\;.
\end{equation}
The amplitude to find three particles in well 2 again contains oscillatory terms, the term
which becomes the leading-order term for large $t$ can be obtained by a calculation analogously to the half-integer resonance in appendix~\ref{sec:second}
\begin{eqnarray}
\left(a_3^{(3)}\right)_{\rm linear} = -\frac{3\Omega^3}4\sum_{k}\sum_{\ell}J_k(2\mu_1/\omega)\\\nonumber \times J_{\ell}(2\mu_1/\omega)J_{1-\ell-k}(2\mu_1/\omega)\frac
t{\sigma_k(\sigma_k+\widetilde{\sigma}_{\ell})}\;.
\end{eqnarray}
This one-third photon resonance can indeed be observed in the
numerics (see Fig.~\ref{fig:drittel}). As this resonance only occurs in third order
perturbation theory (rather than second order for the half-integer resonances), 
the amplitudes would be rather small for interactions as in Fig.~\ref{fig:einhalb}. 
However, choosing an also realistic value of $N\kappa/\Omega = 1.5$ leads to a time-averaged 
population imbalance with a peak-height of the same order of magnitude as in Fig~\ref{fig:einhalb}. 
In a similar manner, smaller fractions could be treated in higher order perturbation theory. As 
the resonances thus are a higher order effect, they will tend to decrease.

\section{Conclusion}
Contrary to the integer-photon peaks~\cite{EckardtEtAl05}, fractional-integer photon peaks
cannot be explained by simply replacing the time-dependent Hamiltonian by a time-independent
Hamiltonian with renormalized tunneling frequencies. As half-integer resonances already appear
for two particles in a double well, this experimentally relevant case~\cite{CheinetEtAl08}
was investigated both numerically and analytically. The perturbation calculations can explain for which parameters the non-integer resonances occur. As the fractional-integer
resonances are only visible for finite
interactions between the particles, they allow to investigate beyond single-particle
effects for very small particle numbers. Experiments similar to Ref.~\cite{CheinetEtAl08}
could thus verify fractional-integer peaks in photon assisted tunneling and thus help to
understand the emergence of effects similar to the non-linearities of a mean-field approach
well below the limit $N\to\infty$.

\acknowledgments

We thank M.~Holthaus for his
continuous support. CW thanks  A.~Eckardt and  A.~L.~Fetter  for insightful discussions; NT and ME acknowledge funding by the Studienstiftung des deutschen Volkes.

\begin{appendix}

\section{\label{sec:second}Second order perturbation theory}
When solving the integral
\begin{equation}
I_{k,\ell}\equiv\int_0^t\frac1{i\sigma_k}\left[\exp\left(i(\sigma_{k}+ \widetilde{\sigma}_{\ell})t'\right)-\exp\left(i\widetilde{\sigma}_{\ell}t'\right)\right]dt'
\end{equation}
 in Eq.~(\ref{eq:textsecond}), one can again assume $\sigma_{k}\ne
0$ and $\widetilde{\sigma}_{\ell}\ne 0$ as  
$\sigma_k=0$ and $\widetilde{\sigma_{\ell}}=0$ would correspond to the
inter-photon resonances discarded here. It then remains to distinguish cases with
\begin{equation}
\label{eq:appsumme}
\sigma_{k}+ \widetilde{\sigma}_{\ell}= 0\;,
\end{equation}
which turn out to be the origin of the half-integer resonance, from those for which this equation is not fulfilled.
If Eq.~(\ref{eq:appsumme}) is fulfilled, one has
\begin{equation}
\label{eq:full}
I_{k,\ell} = \frac1{i\sigma_k}\left[t+\frac1{i{\sigma}_{k}}\left(\exp(-i{\sigma}_{k}t)-1\right)\right]
\end{equation}
otherwise
\begin{equation}
I_{k,\ell} = \frac1{i\sigma_k}\left[\frac{\exp[i({\sigma}_k+\widetilde{\sigma}_{\ell})t]-1}{i({\sigma}_k+\widetilde{\sigma}_{\ell})}-\frac{\exp(i\widetilde{\sigma}_{\ell}t)-1}{i\widetilde{\sigma}_{\ell}}\right]\;.
\end{equation}
Collecting all terms given by Eq.~(\ref{eq:full}), one has the leading-order contribution:
\begin{eqnarray}
\label{eq:leadingorder}
 \left(a_2^{(2)}(t)\right)_{\rm leading-order} =\\\nonumber -\frac{\Omega^2}2
\sum_{k=-\infty}^{\infty}J_k\left(x\right)J_{1-k}\left(x\right)\frac t{\sigma_k}\\\nonumber -\frac{\Omega^2}2
\sum_{k=-\infty}^{\infty}J_k\left(x\right)J_{1-k}\left(x\right)\frac
{\exp(-i{\sigma}_{k}t)-1}{i\sigma_k}\;,
\end{eqnarray}
with
\begin{equation}
x\equiv 2\mu_1/\omega\;,
\end{equation}
and an oscillatory part
\begin{eqnarray}
\label{eq:a2oscil}
 \left(a_2^{(2)}(t)\right)_{\rm oscil} = \quad\quad\\ \nonumber
-\frac{\Omega^2}2\sum_{k=-\infty}^{\infty}\sum_{\ell\ne
  1-k}\frac{J_k(x)J_{\ell}(x)i^{k+l-1}}{\sigma_k}\\ \nonumber
\times\left(\frac{\exp(i(\sigma_k+\widetilde{\sigma}_{\ell})t)-1}{(i(\sigma_k+\widetilde{\sigma}_{\ell}))}-\frac{\exp(i\widetilde{\sigma}_{\ell}t)-1}{i\widetilde{\sigma}_{\ell}}\right)
\;.
\end{eqnarray}

While Eq.~(\ref{eq:leadingorder}) includes the leading-order behavior for large times and
most parameters, it vanishes in the limit~$\mu_1\to 0$. Thus to evaluate the analytic formula
with the help of a computer algebra program, we include the only non-vanishing term for
$\mu_1=0$ to obtain the data displayed in Fig.~\ref{fig:a2}
\begin{eqnarray}
\label{eq:a22nah}
 \left(a_2^{(2)}(t)\right)_{\rm approx} =  \left(a_2^{(2)}(t)\right)_{\rm leading-order} \quad\\ \nonumber
-\frac{\Omega^2}2\frac{J_0(x)J_{0}(x)i^{-1}}{\sigma_0}\quad\quad\\ \nonumber
\times\left(\frac{\exp(i(\sigma_0+\widetilde{\sigma}_{0})t)-1}{(i(\sigma_0+\widetilde{\sigma}_{0}))}-\frac{\exp(i\widetilde{\sigma}_{0}t)-1}{i\widetilde{\sigma}_{0}}\right)
\;.
\end{eqnarray}
\end{appendix}


\begin{thebibliography}{32}
\expandafter\ifx\csname natexlab\endcsname\relax\def\natexlab#1{#1}\fi
\expandafter\ifx\csname bibnamefont\endcsname\relax
  \def\bibnamefont#1{#1}\fi
\expandafter\ifx\csname bibfnamefont\endcsname\relax
  \def\bibfnamefont#1{#1}\fi
\expandafter\ifx\csname citenamefont\endcsname\relax
  \def\citenamefont#1{#1}\fi
\expandafter\ifx\csname url\endcsname\relax
  \def\url#1{\texttt{#1}}\fi
\expandafter\ifx\csname urlprefix\endcsname\relax\def\urlprefix{URL }\fi
\providecommand{\bibinfo}[2]{#2}
\providecommand{\eprint}[2][]{\url{#2}}

\bibitem[{\citenamefont{Grossmann et~al.}(1991)\citenamefont{Grossmann,
  Dittrich, Jung, and {{H\"anggi}}}}]{GrossmannEtAl91}
\bibinfo{author}{\bibfnamefont{F.}~\bibnamefont{Grossmann}},
  \bibinfo{author}{\bibfnamefont{T.}~\bibnamefont{Dittrich}},
  \bibinfo{author}{\bibfnamefont{P.}~\bibnamefont{Jung}}, \bibnamefont{and}
  \bibinfo{author}{\bibfnamefont{P.}~\bibnamefont{{{H\"anggi}}}},
  \bibinfo{journal}{Phys.\ Rev.\ Lett.} \textbf{\bibinfo{volume}{67}},
  \bibinfo{pages}{516} (\bibinfo{year}{1991}).

\bibitem[{\citenamefont{Holthaus}(1992)}]{Holthaus92}
\bibinfo{author}{\bibfnamefont{M.}~\bibnamefont{Holthaus}},
  \bibinfo{journal}{Phys.\ Rev.\ Lett.} \textbf{\bibinfo{volume}{69}},
  \bibinfo{pages}{1596} (\bibinfo{year}{1992}).

\bibitem[{\citenamefont{Eckardt et~al.}(2005)\citenamefont{Eckardt,
  Jinasundera, Weiss, and Holthaus}}]{EckardtEtAl05}
\bibinfo{author}{\bibfnamefont{A.}~\bibnamefont{Eckardt}},
  \bibinfo{author}{\bibfnamefont{T.}~\bibnamefont{Jinasundera}},
  \bibinfo{author}{\bibfnamefont{C.}~\bibnamefont{Weiss}}, \bibnamefont{and}
  \bibinfo{author}{\bibfnamefont{M.}~\bibnamefont{Holthaus}},
  \bibinfo{journal}{Phys.\ Rev.\ Lett.} \textbf{\bibinfo{volume}{95}},
  \bibinfo{pages}{200401} (\bibinfo{year}{2005}).

\bibitem[{\citenamefont{Creffield and Sols}(2008)}]{CreffieldSols08}
\bibinfo{author}{\bibfnamefont{C.~E.} \bibnamefont{Creffield}}
  \bibnamefont{and} \bibinfo{author}{\bibfnamefont{F.}~\bibnamefont{Sols}},
  \bibinfo{journal}{Phys.\ Rev.\ Lett.} \textbf{\bibinfo{volume}{100}},
  \bibinfo{eid}{250402} (\bibinfo{year}{2008}).

\bibitem[{\citenamefont{Kierig et~al.}(2008)\citenamefont{Kierig,
  Schnorrberger, Schietinger, Tomkovic, and Oberthaler}}]{KierigEtAl08}
\bibinfo{author}{\bibfnamefont{E.}~\bibnamefont{Kierig}},
  \bibinfo{author}{\bibfnamefont{U.}~\bibnamefont{Schnorrberger}},
  \bibinfo{author}{\bibfnamefont{A.}~\bibnamefont{Schietinger}},
  \bibinfo{author}{\bibfnamefont{J.}~\bibnamefont{Tomkovic}}, \bibnamefont{and}
  \bibinfo{author}{\bibfnamefont{M.~K.} \bibnamefont{Oberthaler}},
  \bibinfo{journal}{Phys.\ Rev.\ Lett.} \textbf{\bibinfo{volume}{100}},
  \bibinfo{eid}{190405} (\bibinfo{year}{2008}).

\bibitem[{\citenamefont{Zenesini et~al.}(2009)\citenamefont{Zenesini, Lignier,
  Ciampini, Morsch, and Arimondo}}]{Zenesini09}
\bibinfo{author}{\bibfnamefont{A.}~\bibnamefont{Zenesini}},
  \bibinfo{author}{\bibfnamefont{H.}~\bibnamefont{Lignier}},
  \bibinfo{author}{\bibfnamefont{D.}~\bibnamefont{Ciampini}},
  \bibinfo{author}{\bibfnamefont{O.}~\bibnamefont{Morsch}}, \bibnamefont{and}
  \bibinfo{author}{\bibfnamefont{E.}~\bibnamefont{Arimondo}},
  \bibinfo{journal}{Phys.\ Rev.\ Lett.} \textbf{\bibinfo{volume}{102}},
  \bibinfo{eid}{100403} (\bibinfo{year}{2009}).

\bibitem[{\citenamefont{Kohler and Sols}(2003)}]{KohlerSols03}
\bibinfo{author}{\bibfnamefont{S.}~\bibnamefont{Kohler}} \bibnamefont{and}
  \bibinfo{author}{\bibfnamefont{F.}~\bibnamefont{Sols}}, \bibinfo{journal}{New
  J. Phys.} \textbf{\bibinfo{volume}{5}}, \bibinfo{pages}{94}
  (\bibinfo{year}{2003}).

\bibitem[{\citenamefont{Creffield and Monteiro}(2006)}]{CreffieldMonteiro06}
\bibinfo{author}{\bibfnamefont{C.~E.} \bibnamefont{Creffield}}
  \bibnamefont{and} \bibinfo{author}{\bibfnamefont{T.~S.}
  \bibnamefont{Monteiro}}, \bibinfo{journal}{Phys.\ Rev.\ Lett.}
  \textbf{\bibinfo{volume}{96}}, \bibinfo{pages}{210403}
  (\bibinfo{year}{2006}).

\bibitem[{\citenamefont{Sias et~al.}(2008)\citenamefont{Sias, Lignier, Singh,
  Zenesini, Ciampini, Morsch, and Arimondo}}]{SiasEtAl08}
\bibinfo{author}{\bibfnamefont{C.}~\bibnamefont{Sias}},
  \bibinfo{author}{\bibfnamefont{H.}~\bibnamefont{Lignier}},
  \bibinfo{author}{\bibfnamefont{Y.~P.} \bibnamefont{Singh}},
  \bibinfo{author}{\bibfnamefont{A.}~\bibnamefont{Zenesini}},
  \bibinfo{author}{\bibfnamefont{D.}~\bibnamefont{Ciampini}},
  \bibinfo{author}{\bibfnamefont{O.}~\bibnamefont{Morsch}}, \bibnamefont{and}
  \bibinfo{author}{\bibfnamefont{E.}~\bibnamefont{Arimondo}},
  \bibinfo{journal}{Phys.\ Rev.\ Lett.} \textbf{\bibinfo{volume}{100}},
  \bibinfo{pages}{040404} (\bibinfo{year}{2008}).

\bibitem[{\citenamefont{Shapiro}(1963)}]{Shapiro63}
\bibinfo{author}{\bibfnamefont{S.}~\bibnamefont{Shapiro}},
  \bibinfo{journal}{Phys. Rev. Lett.} \textbf{\bibinfo{volume}{11}},
  \bibinfo{pages}{80} (\bibinfo{year}{1963}).

\bibitem[{\citenamefont{Albiez et~al.}(2005)\citenamefont{Albiez, Gati,
  Folling, Hunsmann, Cristiani, and Oberthaler}}]{AlbiezEtAl05}
\bibinfo{author}{\bibfnamefont{M.}~\bibnamefont{Albiez}},
  \bibinfo{author}{\bibfnamefont{R.}~\bibnamefont{Gati}},
  \bibinfo{author}{\bibfnamefont{J.}~\bibnamefont{Folling}},
  \bibinfo{author}{\bibfnamefont{S.}~\bibnamefont{Hunsmann}},
  \bibinfo{author}{\bibfnamefont{M.}~\bibnamefont{Cristiani}},
  \bibnamefont{and} \bibinfo{author}{\bibfnamefont{M.~K.}
  \bibnamefont{Oberthaler}}, \bibinfo{journal}{Phys.\ Rev.\ Lett.}
  \textbf{\bibinfo{volume}{95}}, \bibinfo{pages}{010402}
  (\bibinfo{year}{2005}).

\bibitem[{\citenamefont{Cheinet et~al.}(2008)\citenamefont{Cheinet, Trotzky,
  Feld, Schnorrberger, Moreno-Cardoner, F\"{o}lling, and
  Bloch}}]{CheinetEtAl08}
\bibinfo{author}{\bibfnamefont{P.}~\bibnamefont{Cheinet}},
  \bibinfo{author}{\bibfnamefont{S.}~\bibnamefont{Trotzky}},
  \bibinfo{author}{\bibfnamefont{M.}~\bibnamefont{Feld}},
  \bibinfo{author}{\bibfnamefont{U.}~\bibnamefont{Schnorrberger}},
  \bibinfo{author}{\bibfnamefont{M.}~\bibnamefont{Moreno-Cardoner}},
  \bibinfo{author}{\bibfnamefont{S.}~\bibnamefont{F\"{o}lling}},
  \bibnamefont{and} \bibinfo{author}{\bibfnamefont{I.}~\bibnamefont{Bloch}},
  \bibinfo{journal}{Phys.\ Rev.\ Lett.} \textbf{\bibinfo{volume}{101}},
  \bibinfo{eid}{090404} (\bibinfo{year}{2008}).

\bibitem[{\citenamefont{Shirley}(1965)}]{Shirley65}
\bibinfo{author}{\bibfnamefont{J.~H.} \bibnamefont{Shirley}},
  \bibinfo{journal}{Phys. Rev.} \textbf{\bibinfo{volume}{138}},
  \bibinfo{pages}{B979} (\bibinfo{year}{1965}).

\bibitem[{\citenamefont{Jinasundera et~al.}(2006)\citenamefont{Jinasundera,
  Weiss, and Holthaus}}]{JinasunderaEtAl06}
\bibinfo{author}{\bibfnamefont{T.}~\bibnamefont{Jinasundera}},
  \bibinfo{author}{\bibfnamefont{C.}~\bibnamefont{Weiss}}, \bibnamefont{and}
  \bibinfo{author}{\bibfnamefont{M.}~\bibnamefont{Holthaus}},
  \bibinfo{journal}{Chem. Phys.} \textbf{\bibinfo{volume}{322}},
  \bibinfo{pages}{118} (\bibinfo{year}{2006}).

\bibitem[{\citenamefont{Eckardt and Holthaus}(2008)}]{EckardtHolthaus08b}
\bibinfo{author}{\bibfnamefont{A.}~\bibnamefont{Eckardt}} \bibnamefont{and}
  \bibinfo{author}{\bibfnamefont{M.}~\bibnamefont{Holthaus}},
  \bibinfo{journal}{Phys.\ Rev.\ Lett.} \textbf{\bibinfo{volume}{101}},
  \bibinfo{eid}{245302} (\bibinfo{year}{2008}).

\bibitem[{\citenamefont{Strzys et~al.}(2008)\citenamefont{Strzys, Graefe, and
  Korsch}}]{StrzysEtAl08}
\bibinfo{author}{\bibfnamefont{M.~P.} \bibnamefont{Strzys}},
  \bibinfo{author}{\bibfnamefont{E.~M.} \bibnamefont{Graefe}},
  \bibnamefont{and} \bibinfo{author}{\bibfnamefont{H.~J.}
  \bibnamefont{Korsch}}, \bibinfo{journal}{New J. Phys.}
  \textbf{\bibinfo{volume}{10}}, \bibinfo{pages}{013024}
  (\bibinfo{year}{2008}).

\bibitem[{\citenamefont{Hai et~al.}({2008})\citenamefont{Hai, Lee, and
  Zhu}}]{HaiEtAl08}
\bibinfo{author}{\bibfnamefont{W.}~\bibnamefont{Hai}},
  \bibinfo{author}{\bibfnamefont{C.}~\bibnamefont{Lee}}, \bibnamefont{and}
  \bibinfo{author}{\bibfnamefont{Q.}~\bibnamefont{Zhu}}, \bibinfo{journal}{{J.
  Phys. B}} \textbf{\bibinfo{volume}{{41}}}, \bibinfo{pages}{{095301}}
  (\bibinfo{year}{{2008}}).

\bibitem[{\citenamefont{Weiss and Jinasundera}(2005)}]{WeissJinasundera05}
\bibinfo{author}{\bibfnamefont{C.}~\bibnamefont{Weiss}} \bibnamefont{and}
  \bibinfo{author}{\bibfnamefont{T.}~\bibnamefont{Jinasundera}},
  \bibinfo{journal}{Phys.\ Rev.\ A} \textbf{\bibinfo{volume}{72}},
  \bibinfo{pages}{053626} (\bibinfo{year}{2005}).

\bibitem[{\citenamefont{Kalosakas et~al.}(2003)\citenamefont{Kalosakas, Bishop,
  and Kenkre}}]{KalosakasEtAl03}
\bibinfo{author}{\bibfnamefont{G.}~\bibnamefont{Kalosakas}},
  \bibinfo{author}{\bibfnamefont{A.~R.} \bibnamefont{Bishop}},
  \bibnamefont{and} \bibinfo{author}{\bibfnamefont{V.~M.}
  \bibnamefont{Kenkre}}, \bibinfo{journal}{Phys. Rev. A}
  \textbf{\bibinfo{volume}{68}}, \bibinfo{pages}{023602}
  (\bibinfo{year}{2003}).

\bibitem[{\citenamefont{Smerzi et~al.}(1997)\citenamefont{Smerzi, Fantoni,
  Giovanazzi, and Shenoy}}]{SmerziEtAl97}
\bibinfo{author}{\bibfnamefont{A.}~\bibnamefont{Smerzi}},
  \bibinfo{author}{\bibfnamefont{S.}~\bibnamefont{Fantoni}},
  \bibinfo{author}{\bibfnamefont{S.}~\bibnamefont{Giovanazzi}},
  \bibnamefont{and} \bibinfo{author}{\bibfnamefont{S.~R.}
  \bibnamefont{Shenoy}}, \bibinfo{journal}{Phys.\ Rev.\ Lett.}
  \textbf{\bibinfo{volume}{79}}, \bibinfo{pages}{4950} (\bibinfo{year}{1997}).

\bibitem[{\citenamefont{Castin and Dalibard}(1997)}]{CastinDalibard97}
\bibinfo{author}{\bibfnamefont{Y.}~\bibnamefont{Castin}} \bibnamefont{and}
  \bibinfo{author}{\bibfnamefont{J.}~\bibnamefont{Dalibard}},
  \bibinfo{journal}{Phys. Rev. A} \textbf{\bibinfo{volume}{55}},
  \bibinfo{pages}{4330} (\bibinfo{year}{1997}).

\bibitem[{\citenamefont{Lesanovsky et~al.}(2006)\citenamefont{Lesanovsky,
  Hofferberth, Schmiedmayer, and Schmelcher}}]{LesanovskyEtAl06}
\bibinfo{author}{\bibfnamefont{I.}~\bibnamefont{Lesanovsky}},
  \bibinfo{author}{\bibfnamefont{S.}~\bibnamefont{Hofferberth}},
  \bibinfo{author}{\bibfnamefont{J.}~\bibnamefont{Schmiedmayer}},
  \bibnamefont{and}
  \bibinfo{author}{\bibfnamefont{P.}~\bibnamefont{Schmelcher}},
  \bibinfo{journal}{Phys. Rev. A} \textbf{\bibinfo{volume}{74}},
  \bibinfo{eid}{033619} (\bibinfo{year}{2006}).

\bibitem[{\citenamefont{Piazza et~al.}(2008)\citenamefont{Piazza, Pezz\'{e},
  and Smerzi}}]{PiazzaEtAl2008}
\bibinfo{author}{\bibfnamefont{F.}~\bibnamefont{Piazza}},
  \bibinfo{author}{\bibfnamefont{L.}~\bibnamefont{Pezz\'{e}}},
  \bibnamefont{and} \bibinfo{author}{\bibfnamefont{A.}~\bibnamefont{Smerzi}},
  \bibinfo{journal}{Phys.\ Rev.\ A} \textbf{\bibinfo{volume}{78}},
  \bibinfo{pages}{051601(R)} (\bibinfo{year}{2008}).

\bibitem[{\citenamefont{Lee et~al.}(2008)\citenamefont{Lee, Fu, and
  Kivshar}}]{LeeEtAl08}
\bibinfo{author}{\bibfnamefont{C.}~\bibnamefont{Lee}},
  \bibinfo{author}{\bibfnamefont{L.-B.} \bibnamefont{Fu}}, \bibnamefont{and}
  \bibinfo{author}{\bibfnamefont{Y.~S.} \bibnamefont{Kivshar}},
  \bibinfo{journal}{EPL} \textbf{\bibinfo{volume}{81}}, \bibinfo{pages}{60006}
  (\bibinfo{year}{2008}).

\bibitem[{\citenamefont{Esteve et~al.}({2008})\citenamefont{Esteve, Gross,
  Weller, Giovanazzi, and Oberthaler}}]{EsteveEtAl08}
\bibinfo{author}{\bibfnamefont{J.}~\bibnamefont{Esteve}},
  \bibinfo{author}{\bibfnamefont{C.}~\bibnamefont{Gross}},
  \bibinfo{author}{\bibfnamefont{A.}~\bibnamefont{Weller}},
  \bibinfo{author}{\bibfnamefont{S.}~\bibnamefont{Giovanazzi}},
  \bibnamefont{and} \bibinfo{author}{\bibfnamefont{M.~K.}
  \bibnamefont{Oberthaler}}, \bibinfo{journal}{{Nature}}
  \textbf{\bibinfo{volume}{{455}}}, \bibinfo{pages}{{1216}}
  (\bibinfo{year}{{2008}}).

\bibitem[{\citenamefont{Yukalov and Yukalova}(2009)}]{YukalovYukalova09}
\bibinfo{author}{\bibfnamefont{V.~I.} \bibnamefont{Yukalov}} \bibnamefont{and}
  \bibinfo{author}{\bibfnamefont{E.~P.} \bibnamefont{Yukalova}},
  \bibinfo{journal}{Laser Phys.\ Lett.} \textbf{\bibinfo{volume}{6}},
  \bibinfo{pages}{235} (\bibinfo{year}{2009}).

\bibitem[{\citenamefont{Lipkin et~al.}(1965)\citenamefont{Lipkin, Meshkov, and
  Glick}}]{LipkinEtAl65}
\bibinfo{author}{\bibfnamefont{H.~J.} \bibnamefont{Lipkin}},
  \bibinfo{author}{\bibfnamefont{N.}~\bibnamefont{Meshkov}}, \bibnamefont{and}
  \bibinfo{author}{\bibfnamefont{A.~J.} \bibnamefont{Glick}},
  \bibinfo{journal}{Nucl. Phys.} \textbf{\bibinfo{volume}{62}},
  \bibinfo{pages}{188} (\bibinfo{year}{1965}).

\bibitem[{\citenamefont{Milburn et~al.}(1997)\citenamefont{Milburn, Corney,
  Wright, and Walls}}]{MilburnEtAl97}
\bibinfo{author}{\bibfnamefont{G.~J.} \bibnamefont{Milburn}},
  \bibinfo{author}{\bibfnamefont{J.}~\bibnamefont{Corney}},
  \bibinfo{author}{\bibfnamefont{E.~M.} \bibnamefont{Wright}},
  \bibnamefont{and} \bibinfo{author}{\bibfnamefont{D.~F.} \bibnamefont{Walls}},
  \bibinfo{journal}{Phys.\ Rev.\ A} \textbf{\bibinfo{volume}{55}},
  \bibinfo{pages}{4318} (\bibinfo{year}{1997}).

\bibitem[{\citenamefont{Weiss and Teichmann}(2008)}]{WeissTeichmann08}
\bibinfo{author}{\bibfnamefont{C.}~\bibnamefont{Weiss}} \bibnamefont{and}
  \bibinfo{author}{\bibfnamefont{N.}~\bibnamefont{Teichmann}},
  \bibinfo{journal}{Phys.\ Rev.\ Lett.} \textbf{\bibinfo{volume}{100}},
  \bibinfo{eid}{140408} (\bibinfo{year}{2008}).

\bibitem[{\citenamefont{Abramowitz and Stegun}(1984)}]{Abramowitz84}
\bibinfo{author}{\bibfnamefont{M.}~\bibnamefont{Abramowitz}} \bibnamefont{and}
  \bibinfo{author}{\bibfnamefont{I.~A.} \bibnamefont{Stegun}},
  \emph{\bibinfo{title}{Pocketbook of Mathematical Functions}}
  (\bibinfo{publisher}{Verlag Harri Deutsch}, \bibinfo{address}{Thun},
  \bibinfo{year}{1984}).

\bibitem[{\citenamefont{Landau and Lifshitz}(2000)}]{Landau00}
\bibinfo{author}{\bibfnamefont{L.~D.} \bibnamefont{Landau}} \bibnamefont{and}
  \bibinfo{author}{\bibfnamefont{E.~M.} \bibnamefont{Lifshitz}},
  \emph{\bibinfo{title}{Course of theoretical physics, Vol. 3}}
  (\bibinfo{publisher}{Butterworth-Heinemann}, \bibinfo{address}{Oxford},
  \bibinfo{year}{2000}).

\bibitem[{\citenamefont{Weiss and Breuer}(2009)}]{WeissBreuer09}
\bibinfo{author}{\bibfnamefont{C.}~\bibnamefont{Weiss}} \bibnamefont{and}
  \bibinfo{author}{\bibfnamefont{H.-P.} \bibnamefont{Breuer}},
  \bibinfo{journal}{Phys. Rev. A} \textbf{\bibinfo{volume}{79}},
  \bibinfo{eid}{023608} (\bibinfo{year}{2009}).

\end{thebibliography}

\end{document}